\documentclass[reprint,aps,prd]{revtex4-1} 
\usepackage{graphicx}
\begin{document}


\title{A large disparity in cosmic reference frames determined from the sky distributions of radio sources and the microwave background radiation}

\author{Ashok K. Singal}
\email{ashokkumar.singal@gmail.com}
\affiliation{Astronomy and Astrophysics Division, Physical Research Laboratory,
Navrangpura, Ahmedabad - 380 009, India }
\date{\today}
\begin{abstract}
The angular distribution of the Cosmic Microwave Background Radiation (CMBR) in sky shows   
a dipole asymmetry, ascribed to the observer's motion (peculiar velocity of the solar system!), relative to the local comoving coordinates. The peculiar velocity thus determined turns out to be $370$ km s$^{-1}$ in the direction RA$=168^{\circ}$, Dec$=-7^{\circ}$. On the other hand, a dipole asymmetry in the sky distribution of radio sources in the NRAO VLA Sky Survey (NVSS) catalog, comprising 1.8 million sources, yielded a value for the observer's velocity to be $\sim 4$ times larger than the CMBR value, though the direction turned out to be in agreement with that of the CMBR dipole. This large difference in observer's speeds with respect to the reference frames of NVSS radio sources and of CMBR, confirmed since by many independent groups, is rather disconcerting as the observer's motion with respect to local comoving coordinates should be independent of  the technique used to determine it. A genuine difference in relative speeds of two cosmic reference frames could jeopardize the cosmological principle, thence it is crucial to confirm such discrepancies using independent samples of radio sources. We here investigate the dipole in the sky distribution of radio sources in the recent TIFR GMRT Sky Survey (TGSS) dataset, comprising 0.62 million sources, to determine  observer's motion. We find a significant disparity in observer's speeds relative to all three reference frames, determined from the radio source datasets and the CMBR, which does not fit with the cosmological principle, a starting point for the standard modern cosmology.
\end{abstract}
\maketitle
\section{INTRODUCTION}
Due to the assumed isotropy of the Universe -- \`a la cosmological principle -- an observer stationary with respect to the comoving coordinates of the cosmic fluid, should find the number counts of distant radio sources as well as the sky brightness therefrom (i.e., an integrated emission from discrete sources per unit solid angle), to be uniform over the sky. However, an observer moving with a velocity $v$ relative to the cosmic fluid will find, as a combined effect of aberration and Doppler boosting, the number counts and the sky brightness to vary as $\propto \delta^{2+x(1+\alpha)}$, where $\delta$ ($=1+(v/c)\cos\theta$, for a non-relativistic case) is the Doppler factor, $c$ is the velocity of light, $\alpha$ ($\approx 0.8$) is the spectral index, defined by $S \propto \nu^{-\alpha}$, and $x$ is the index of the integral source counts of extragalactic radio source population, which follows a power law $N(>S)\propto S^{-x}$ ($x \sim 1$) \cite{6,20,7}.  
The angular variation of the number counts as well as of the sky brightness can be expressed as $1+{\cal D}\cos\theta$, implying a dipole anisotropy over the sky with an amplitude  \cite{6,20,3,4,7,10} 
\begin{equation}
\label{eq:1}
{\cal D}=\left[2+{x(1+\alpha)}\right]\frac{v}{c}\;.
\end{equation}
By observing such angular variation over the sky for a sufficiently large dataset of distant radio sources, one can compute the dipole ${\cal D}$ and thereby velocity $v$ of the observer with respect to the comoving coordinates.

Let $ \bf{\hat{r}}_i$ be the angular position of $i^{th}$ source of observed flux density $S_i$ 
with respect to the stationary observer, who 
should find $\Sigma S_i \bf{\hat{r}}_i=0$. However, for a moving observer,  
due to the dipole anisotropy over the sky, it would yield a finite vector along the direction of the dipole. Let $\bf{\hat{d}}$ be a unit vector in the direction of the dipole, then writing $\Delta {\cal F}=\Sigma S_i\: \bf{\hat{d}}\cdot\bf{\hat{r}}_i$ and ${\cal F}=\Sigma S_i\: |\bf{\hat{d}}\cdot\bf{\hat{r}}_i|$, a summation  over the whole sky determines  magnitude of the dipole in the sky brightness as \cite{20,7,10} 
\begin{equation}
\label{eq:2}
{\cal D}=\frac {{\cal D}_{\rm o}}{k}=\frac {3}{2k}\frac{\Delta {\cal F}}{ {\cal F}}
=\frac {3}{2k}\frac{\Sigma  S_i\: \cos \theta_i}{\Sigma  S_i\: |\cos \theta_i|}\:,
\end{equation}
where $\theta_i$ is the polar angle of the $i^{th}$ source with respect to the dipole direction. 
Here ${\cal D}_{\rm o}$ is the dipole determined from observational data and might be affected by any gaps in the sky coverage and some other factors as discussed later, and $k$ is the correction factor, of the order of unity, and as such would need to be determined numerically for individual samples. 

A study of the angular variation in the temperature distribution of the CMBR has given quite accurate 
measurements of a dipole anisotropy, supposedly arising from the observer's motion (peculiar velocity of the solar system!) of $370$ km s$^{-1}$, in the direction 
$l=264^{\circ}, b=48^{\circ}$ or equivalently, RA$=168^{\circ}$, Dec$=-7^{\circ}$ \cite{1,2,16}. Some earlier attempts to determine the dipole from the observed angular asymmetry in the sky distribution of distant radio sources claimed the radio source dipole to match the CMBR dipole within statistical uncertainties \cite{20,3}.  However, Singal \cite{7}, from a study of the anisotropy in the number counts as well as in the sky brightness from discrete radio sources in the NVSS catalog \cite{5}, covering whole sky north of declination $-40^{\circ}$ and containing $\sim 1.8$ million sources with a flux-density limit $S>3$ mJy  at 1.4 GHz, found the solar peculiar motion  $\sim 4$ times the CMBR value 
at a statistically significant ($\sim 3\sigma$) level. At the same time, the direction of the velocity vector, though, was surprisingly found to be in agreement with the CMBR value. These unexpected findings of the NVSS dipole being many times larger than the CMBR dipole, have since been confirmed in a number of publications \cite{8,9,18,19,12}. Such a 
difference between two dipoles would imply a relative motion between two cosmic reference frames which will be against the cosmological principle on which the whole modern cosmology is based upon. Therefore it is imperative that an investigation of the radio source dipole be made employing some independent radio source samples. 
A recent estimate from the TGSS data \cite{11}, has yielded an even larger amplitude
for the radio dipole \cite{12}, which is all the more disturbing. Here we investigate this radio dipole in further details, by choosing different flux-density levels, to examine the self-consistency of the dipole in the TGSS data and relate the results to those from NVSS catalog by using the spectral index information between the two datasets \cite{17,15}.

\section{TGSS Dataset}
TIFR  GMRT  Sky  Survey (TGSS) is a 150 MHz, continuum  survey, carried out between 2010 and 2012 using the Giant  Metrewave  Radio  Telescope (GMRT) \cite{14}, and  the  raw  data are available  at the GMRT archive. A First Alternative Data Release of the TGSS (TGSS-ADR1) \cite{11}, that includes direction-dependent calibration and imaging, is available online in the public domain. The TGSS-ADR1, henceforth called TGSS, dataset covers whole sky north of declination $-53^{\circ}$, a  total of $3.6\pi$ sr, amounting to $90\%$ of the celestial sphere, with an rms noise  below  5 mJy/beam and an  approximate  resolution  of  $25''\times25''$.  Using  a  detection limit of 7-sigma, the TGSS catalog comprises 0.62 Million radio sources with an accuracy of about $2''$ or better in RA and Dec. From the spectral index data of common sources in the TGSS and NVSS catalogs \cite{12,17,15}, it has been demonstrated that the number counts in the two datasets could be compared above the flux-density limits of 100 mJy and 20 mJy respectively, using a relation $S_{\rm NVSS}\approx 0.2\: S_{\rm TGSS}$, and that the two catalogs are essentially complete above these respective flux-density limits. 

As the TGSS catalog \cite{11} has a gap of sources for Dec $<-53^{\circ}$, in that case our assumption of 
$\Sigma S_i \bf{\hat{r}}_i=0$ for a stationary observer does not hold good. However if we drop all sources with  
Dec $> 53 ^{\circ}$ as well, then with equal and opposite gaps on two opposite 
sides, $\Sigma S_i \bf{\hat{r}}_i=0$ is valid for a stationary observer \cite{7}. 
Exclusion of such sky-strips, which affect the forward and backward measurements 
identically, to a first order do not have systematic effects on the 
results \cite{6}. 
We also exclude all sources from our sample which lie in the galactic plane ($|b|<10^{\circ}$),
otherwise a large number of galactic sources in the galactic plane is likely to have an unwanted influence on our determination of the radio source dipole. 
To ascertain effects of some systematics like local clustering (mainly the Virgo super-cluster), we also examined any alterations in our results by restricting our dataset to regions outside the super-galactic plane by rejecting sources with low super-galactic latitude ($|{\rm SGB}| < 10^{\circ}$).
\begin{table*}
\begin{center}
\caption{The dipole and velocity vector from the sky brightness for the TGSS dataset}
\hskip4pc\vbox{\columnwidth=33pc
\begin{tabular}{ccccccccccccccc}
\tableline\tableline 
 Flux-density Range & $N$ &&  RA && Dec && ${\cal D}_{\rm o}$   && $k$ && ${\cal D}$  &  & $v$\\
 (mJy) && & ($^{\circ}$)& & ($^{\circ}$) && ($10^{-2}$) &&&& ($10^{-2}$) && ($10^{3}$ km s$^{-1}$) \\ \hline
$5000>S\geq 250$ & 098205 && $174 \pm 11$ && $-08 \pm 10$ &&$6.43\pm 0.73$ &&1.14 &&$5.64\pm 0.64$ && $4.33\pm0.49$ \\
$5000>S\geq 200$ & 122549 && $174 \pm 11$ && $-06 \pm 10$ &&$6.19\pm 0.68$ && 1.14&&$5.43\pm 0.60$ && $4.32\pm0.48$ \\
$5000>S\geq 150$ & 160133 & & $173 \pm 10$ && $-06 \pm 09$ &&$5.98\pm 0.63$ &&1.11 &&$5.39\pm 0.57$ && $4.45\pm0.47$ \\
$5000>S\geq 100$ & 226242& & $172 \pm 10$ && $-05 \pm 09$ &&$5.72\pm 0.58$ &&1.09& &$5.24\pm 0.53$ && $4.47\pm0.45$ \\
\tableline
\end{tabular}
}
\end{center}
\end{table*}
\begin{table*}
\begin{center}
\caption{The dipole and velocity vector for the TGSS dataset with $|{\rm SGB}|>10^\circ$}
\hskip4pc\vbox{\columnwidth=33pc
\begin{tabular}{ccccccccccccccc}
\tableline\tableline 
 Flux-density Range & $N$ &&  RA && Dec && ${\cal D}_{\rm o}$   && $k$ && ${\cal D}$  &  & $v$\\
 (mJy) && & ($^{\circ}$)& & ($^{\circ}$) && ($10^{-2}$) &&&& ($10^{-2}$) && ($10^{3}$ km s$^{-1}$) \\ \hline
$5000>S\geq 250$ & 082336 && $170 \pm 12$ && $-12 \pm 11$ && $6.70\pm 0.82$ &&1.15 & &$5.83\pm 0.71$ && $4.47\pm0.54$ \\
$5000>S\geq 200$ & 102700 && $169 \pm 12$ && $-10 \pm 11$ && $6.41\pm 0.76$ &&1.14 & &$5.62\pm 0.67$ && $4.47\pm0.53$ \\
$5000>S\geq 150$ & 134106  && $167 \pm 11$ && $-10 \pm 10$ && $6.15\pm 0.71$ &&1.12 & &$5.49\pm 0.63$ && $4.54\pm0.52$ \\
$5000>S\geq 100$ & 189416 && $167 \pm 11$ && $-10 \pm 10$ && $5.87\pm 0.65$ &&1.10 && $5.34\pm 0.59$ && $4.56\pm0.50$ \\
\tableline
\end{tabular}
}
\end{center}
\end{table*}

We used Monte--Carlo simulations to create an artificial radio sky with a similar number density of sources as in the TGSS catalog, distributed at random positions in 
the sky. However, for the flux-density distribution we took the observed TGSS sample, so that the source counts remain unchanged. 
On this we superimposed Doppler boosting and aberration effects of an assumed motion, choosing a different velocity vector for each simulation. 
This artificial sky was then used to retrieve the velocity vector under 
conditions similar as in our actual TGSS sample (e.g., with $|{\rm Dec}|> 53 ^{\circ}, |b|<10^{\circ}$ gaps in the sky), and compared with the input value in that particular realization. This not only validated our 
procedure as well as the computer routine, but also helped us make an estimate of errors in the dipole co-ordinates from one hundred simulations we made, each time with randomly chosen radio source sky positions and a different velocity vector assumed for the solar peculiar motion. 
The error in ${\Delta {\cal F}}/{\cal F}$ is given by $(\Sigma S_i^2)^{1/2}/{\sqrt 3}{\cal F}$ \cite{7}, which allows us to compute error in the dipole magnitude $\cal D$ from Eq.~(\ref{eq:2}).

\section{Results and Discussion}
\subsection{Sky brightness}
As a relatively small number of strong sources at high flux-density levels could introduce large statistical 
fluctuations in the sky brightness, we have restricted our sample here to below 5,000 mJy level. At the other end we chose 
100 mJy as the lowest cut-off limit since the TGSS catalog is essentially complete only above that flux-density level \cite{11,17}. 

Results for the dipole, determined from the anisotropy in sky brightness for the TGSS dataset, for sources in various flux-density bins, are presented in Table~I. Here $N$ is the total number of sources in the corresponding flux-density bin, RA and Dec give the dipole direction in sky, $\cal D_{\rm o}$ is the 'raw' dipole value computed from $3\Delta {\cal F}/ 2{\cal F}$ that needs to be corrected for various effects as explained below, $k$ being the correction factor.

The numerical factor $k$ includes the effect of gaps in sky coverage of the data as well as 
the effects, if any, of the variations of power index $x$ and the spectral index $\alpha$ with flux density, and is therefore determined separately for each flux-density bin. 
Now, from Table~I, our estimate of the direction of the dipole (RA$=172^{\circ} \pm 10^{\circ}$, Dec$=-05^{\circ} \pm 09^{\circ}$), is quite in agreement with that determined from the CMBR 
(RA$=168^{\circ}$, Dec$=-7^{\circ}$, with errors less than a degree) \cite{1,2,16}. 
Taking this into account we can make our estimates of $k$ better by using this information in our simulations. Accordingly, these simulations differ from the ones above in the sense that the dipole used now is in the direction of the CMBR dipole direction, superimposed on randomly assigned source positions in the sky. It should be noted that this procedure allows us to make a more realistic assessment of the influence of sky-gaps on our determination of the dipole magnitude and does not introduce bias of any kind. A comparison of the dipole derived from each simulation with the input dipole magnitude yielded the correction for that simulation. A set of 100 different simulations was used to determine an average value of the correction factor $k$, given in Table I.  The corrected dipole $\cal D$ is obtained by dividing the 'raw' dipole strength $\cal D_{\rm o}$ by $k$ for each flux-density bin.

From Table~I we see a trend that the dipole strength ${\cal D}$ falls systematically with a decrease in the lower flux-density cut-off. From 250 mJy to 100 mJy, ${\cal D}$  gradually falls as much as by $\sim 10 \%$, though it comes nowhere near an order of magnitude weaker CMBR dipole. For a given peculiar velocity $v$, the dipole $\cal D$ could get affected by two quantities, the power law index $x$ and the spectral index $\alpha$ (Eq.~(\ref{eq:1})). 
To estimate effects of the power index $x$ on $\cal D$, we have plotted in Fig. (1) the integrated source counts, $N(>S)$ for different $S$ for the TGSS and NVSS samples. The index $x$  in the power law relation, $N(>S)\propto S^{-x}$, can be estimated from the slope of the $\log-\log S$ plot in Fig. (1), where we find that the index $x$ steepens from low to high flux-density levels for both samples. From piece-wise straight line fits to the $\log N-\log S$ data, we find that $x$ steepens from  $-0.85$ ($-0.95$) at 100 (20) mJy to $-1.7$ ($-1.65$) at 5000 (1000) mJy, with a value $-1.05$ ($-1.1$) at 250 (50) mJy in the TGSS (NVSS) data. It should be noted that the value of $x$ that enters into Eq.~(\ref{eq:1}) is the one at the lower cut-off flux density of the bin. On the other hand variation in $\alpha$ is much smaller. From a comparison of TGSS and NVSS samples, the spectral index for the flux-density range $100<S_{\rm TGSS}<200$ mJy was found to be $0.758\pm 0.245$ while for $S_{\rm TGSS}>200$ mJy it turned out to be $0.802\pm 0.225$ \cite{15}. This slight steepening of $\alpha$ with flux density would affect the dipole value only by $\stackrel{<}{_{\sim}} 1 \%$. For definiteness, we use $\alpha=0.76$ for the $100$ ($20$) mJy bin and $\alpha=0.8$ for the $200$ ($40$) mJy bin in the TGSS (NVSS) data. For other flux-density bins we use interpolated or extrapolated values for $x$ and $\alpha$. The peculiar velocity $v$, calculated accordingly from Eq.~(\ref{eq:1}), is listed for each flux-density bin in Table I, where no significant trend with changing flux-density cut-off levels is seen. The underlying assumption throughout is that $v$ represents motion of the observer (solar system!), with respect to the corresponding reference frame of radio sources, in the direction given by RA and Dec of the dipole, and its value should not vary from one flux-density bin to another.

As we mentioned above, the direction of the dipole from TGSS data is quite in agreement with that of the CMBR dipole. However, the  strength of the dipole (${\cal D}\simeq 5.2 \times 10^{-2}$, the best estimate in Table~I) appears an order of magnitude larger than the CMBR dipole ($\simeq 0.47 \times 10^{-2}$) \cite{7}, even though it is smaller by a factor of $\sim 1.4$  than an earlier estimate ($ 7 \times 10^{-2}$) \cite{12}. When compared with the dipole determined from the sky brightness in the NVSS dataset \cite{7}, the TGSS dipole is a factor of $\sim 2.5$ stronger  than the NVSS dipole ($\simeq 2.1 \times 10^{-2}$). 
\begin{table*}
\begin{center}
\caption{The velocity vector from the number counts}
\hskip4pc\vbox{\columnwidth=33pc
\begin{tabular}{ccccccccccccccc}
\tableline\tableline 
 Flux-density cut-off & $N$ &&  RA && Dec &
 & ${\cal D}_{\rm o}$   && $k$ && ${\cal D}$  &  & $v$\\
 (mJy) && & ($^{\circ}$)& & ($^{\circ}$) && ($10^{-2}$) &&&& ($10^{-2}$) && ($10^{3}$ km s$^{-1}$) \\ \hline
$\geq 250$ & 099736 &&  $168\pm 10$ &&  $-08\pm 09$ && $5.27\pm 0.53$ && 1.05& & $5.02\pm 0.50$ && $3.85\pm0.38$ \\
$\geq 200$ & 124080 &&  $166\pm 10$ &&  $-03\pm 09$ && $4.89\pm 0.47$ && 1.05& & $4.65\pm 0.45$ && $3.70\pm0.36$ \\
$\geq 150$ & 161664&&  $164\pm 09$ &&  $-01\pm 08$ && $4.65\pm 0.42$ && 1.00& & $4.65\pm 0.42$ && $3.84\pm0.35$ \\
$\geq 100$ & 227773 &&  $162\pm 09$ &&  $+03\pm 08$ && $4.24\pm 0.35$ && 0.96&& $4.42\pm 0.37$ && $3.77\pm0.32$ \\
\tableline
\end{tabular}
}
\end{center}
\end{table*}
\begin{figure}
\includegraphics[width=\columnwidth]{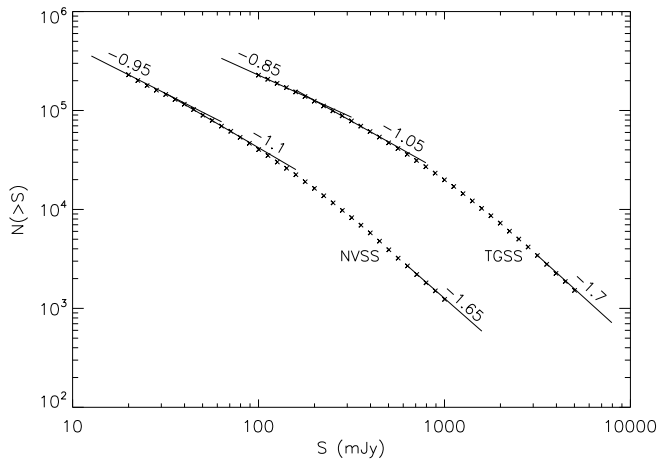}
\caption{A plot of the integrated source counts $N(>S)$ against $S$, for the TGSS and NVSS  samples, showing the power law  behavior ($N(>S)\propto S^{-x}$) of the source counts. From  piece-wise straight line fits to data in different flux-density ranges in either sample, index $x$ appears to steepen for stronger sources, as shown by continuous lines with the best-fit $x$ values shown above.}
\end{figure}

To rule out the possibility that the excessive dipole strength in Table~I might be the result of some local clustering (e.g., the Virgo 
super-cluster), we determined the dipole from the sky brightness from radio sources outside the 
super-galactic plane by dropping sources with low super-galactic latitude, $|{\rm SGB}|<10^\circ$. 
We see that the computed dipole (Table~II) is still an order of magnitude larger than the CMBR dipole, but lies in the same direction, and that this anomalous result is not due to a local clustering. 
\subsection{Number counts} 
We have determined the dipole and the solar peculiar velocity from the number counts as well. First the direction of the dipole was determined from $\Sigma\bf{\hat{r}}_i$. With $\bf{\hat{d}}$ as a unit vector in the direction of the dipole, we define the fractional difference as
\begin{eqnarray}
\label{eq:3}
\frac {\Delta {\cal N}}{{\cal N}}&=& \frac{\Sigma \bf{\hat{d}}\cdot\bf{\hat{r}}_i}{\Sigma |\bf{\hat{d}}\cdot\bf{\hat{r}}_i|}= \frac{\Sigma \cos \theta_i}{\Sigma |\cos \theta_i|},
\end{eqnarray}
where $\theta_i$ is the polar angle of the $i^{th}$ source with respect to the dipole direction. 
The dipole magnitude is then calculated from the fractional difference 
\begin{eqnarray}
\label{eq:4}
{\cal D}=\frac {{\cal D}_{\rm o}}{k}= \frac{3}{2k}\frac {\Delta {\cal N}}{{\cal N}}\:,
\end{eqnarray}
similar to that from ${\Delta {\cal F}}/{\cal F}$ in the case of sky brightness (Eq.~(\ref{eq:2})). Since, unlike in the case of sky brightness, a small number of bright sources do not adversely affect the number counts, in the latter case we have relaxed the upper limit of 5000 mJy on the flux density. 

Like in the case of sky brightness, here too we created an artificial radio sky with sources distributed at random positions in the sky, but with a flux-density distribution as of the TGSS sample, so that the source counts remain unchanged. On this was superimposed a mock dipole oriented in a random direction and of random magnitude. One hundred such independent simulations were made to estimate the expected errors in the dipole co-ordinates. 
The error in ${\Delta {\cal N}}/{\cal N}$ is given by $2/\sqrt {3N}$, then from Eq.~(\ref{eq:4}), error in $\cal D_{\rm o}$ is $\sqrt {3/N}$. For estimating $k$, another set of 100 simulations were made by choosing dipoles oriented in the direction of the CMBR dipole direction, superimposed on randomly assigned source positions in the sky. Such derived correction factor $k$ was used to
divide $\cal D_{\rm o}$ to get $\cal D$ for each flux-density bin. The peculiar velocity $v$, was then calculated using appropriate $x$ nd $\alpha$ values for each flux-density bin.

Results from the number count are summarized in Table~III. 
Comparing with Table~I we notice that the directions of the dipoles determined both from the sky brightness as well as from the number counts, are consistent with that of the CMBR. However, the number counts yield a magnitude of the dipole (and the thereby inferred solar peculiar velocity) to be somewhat smaller ($\sim 15 \%$) than that from the radio source sky brightness. It should be noted that in the sky brightness case, stronger sources get more weight, while in the number counts method, weaker source, being more numerous, dominate the dipole determination, so the results could differ somewhat.
At a first look it may appear that since in the case of sky brightness, the number counts are being weighted by the flux density, for a power law integrated counts, even with a constant $x$, this may result in a dipole estimate to be higher by a factor of $\sim 1.4$ \cite{8}. However, it has been shown \cite{10} that since the flux-density cut-offs for all directions, including the forward and backward directions, are selected in the moving observer's frame and not in the stationary observer's frame, the above argument does not hold good. In any case, the dipole strength in number counts still remains an order of magnitude larger than the CMBR value.

\begin{table*}
\caption{The dipole magnitude and speed estimates for the TGSS and NVSS samples with respect to the CMBR dipole direction}
\hskip4pc\vbox{\columnwidth=33pc
\begin{tabular}{ccccccccccccccccc}
\tableline\tableline 
 sample& $\nu$& $S$  && $N$ &  $\sigma_{\rm N}$ &$N_1$ & $N_2$ & $\delta N$ & $\delta N/ \sigma_N$ &   ${\cal D}_{\rm o}$   && $k$ &&  $\cal D$&  & $v$ \\
& (MHz)& (mJy) && &($\sqrt N$) &  && ($N_1-N_2$)&&($ 10^{-2}$)  &&& &($ 10^{-2}$)  && ($10^{3}$ km s$^{-1}$)
  \\ \hline
TGSS& 150&$>250$ &  & 99736 & 316 & 51254 & 48482 & 2772 & 8.8& $5.56\pm 0.63$ && 1.14&& $4.87\pm 0.56$ & &3.74 $\pm 0.43$ \\
TGSS& 150&$>200$  & & 124080 & 352 & 63648 & 60432 & 3216 & 9.1&  $5.18\pm 0.57$ && 1.15&&$4.51\pm 0.49$ & &3.59 $\pm 0.39$ \\
TGSS& 150&$>150$ & & 161664 & 402 & 82852 & 78812 & 4040 & 10.0&  $5.00\pm 0.50$ && 1.09&&$4.59\pm 0.46$ & &3.79 $\pm 0.38$ \\
TGSS& 150& $>100$ & & 227773 & 477 & 116532 & 111241 & 5291 & 11.1& $4.65\pm 0.42$ && 1.05&& $4.42\pm 0.40$ && 3.77 $\pm 0.34$ \\
\\

NVSS& 1400&$>50$ & & 91652 & 303  & 46372 &  45280 & 1092  & 3.6& $2.38\pm 0.66$ && 1.29& &$1.85\pm 0.51$ && 1.39 $\pm 0.38$ \\
NVSS& 1400&$>40$ & & 115905 & 340  & 58547 & 57358 & 1189  & 3.5& $2.05\pm 0.59$ && 1.25& &$1.64\pm 0.47$ && 1.26 $\pm 0.36$ \\
NVSS& 1400&$>30$ & & 155110 & 394  & 78434 & 76676 & 1758  & 4.5&  $2.27\pm 0.51$ && 1.21&&$1.87\pm 0.42$ && 1.48 $\pm 0.33$ \\
NVSS& 1400&$>20$ & & 229551 & 479 & 115932 & 113619 & 2313  & 4.8& $2.02\pm 0.42$ && 1.15&& $1.75\pm 0.36$ && 1.43 $\pm 0.29$ \\
\tableline
\end{tabular}
}
\end{table*}

If we now compare the dipole determined from number counts for the TGSS dataset with that determined from the NVSS dataset \cite{7,18,8,9,19,12}, we find that the directions of the dipole from both these datasets match well with the CMBR measurements, implying  
that the cause of the dipoles is common and a peculiar motion of the solar system seems to be the only reasonable interpretation for 
that. However such a statistically significant disparity in their 
magnitudes, with the TGSS dipole being an order of magnitude (a factor of $\sim 10$) larger than the CMBR dipole, while the NVSS dipole being $\sim 4$ times larger than the CMBR dipole, is rather unsettling. 
\subsection{Radio survey dipoles with respect to the CMBR dipole direction} 
The fact that the directions of the dipole from the radio source data and the CMBR measurements are matching well, suggests 
that the direction of the CMBR dipole, known with high accuracy, could be taken to be the  direction for the radio source dipoles too. However, we need to first explicitly examine for both TGSS and NVSS datasets if there exist indeed dipoles in the radio source sky distribution with respect to the CMBR dipole direction. For this we compute the dipole strength and the inferred velocity for both 
radio source datasets, but now with respect to the CMBR dipole direction, viz. RA$=168^{\circ}$, Dec$=-7^{\circ}$. For this we employ an alternate procedure which is more transparent, simpler in nature and more easily visualized. 

Using the great circle at $90^\circ$ from the CMBR dipole direction, we divide the sky in two equal hemispheres, $\Sigma_1$ and $\Sigma_2$, with $\Sigma_1$ containing the CMBR dipole, and $\Sigma_2$ containing the direction opposite to the CMBR dipole. Then if there is indeed a motion of the observer along the CMBR dipole direction, due to a combined effect of the aberration and Doppler boosting, the number counts will have a dipole anisotropy, $1+{\cal D}\cos\theta$, over the sky with an amplitude ${\cal D}=[2+x(1+\alpha)]\:v/c$ (Eq.~(\ref{eq:1})), $\theta$ being the angle measured from the CMBR dipole direction. Then the number of sources in the hemisphere $\Sigma_1$, should be larger than the number of sources in the hemisphere $\Sigma_2$. Let $\phi$ be a complementary angle to $\theta$, i.e. $\phi=\pi/2 - \theta$, with $\phi$ measured towards the CMBR dipole direction, starting from the great circle that divides the sky into hemispheres $\Sigma_1$ and $\Sigma_2$. Now, counting from the great circle, if we denote by $N_1$ the number of sources between 0 and $\phi$ in $\Sigma_1$,  then we can write 
\begin{equation}
\label{eq5}
N_1=2\pi N_0{\int^{\phi}_{0}(1+ {\cal D}\sin\phi) \cos \phi\:{\rm d}\phi}, 
\end{equation}
where $N_0$ is the number density per unit solid angle for an isotropic distribution, in the absence of any peculiar motion.
Similarly we can write the number of sources in the opposite hemisphere $\Sigma_2$ as
\begin{equation}
\label{eq:6}
N_2=2\pi N_0{\int^{0}_{-\phi}(1+ {\cal D}\sin\phi) \cos \phi\:{\rm d}\phi}, 
\end{equation}

Then the fractional excess in number of sources in sky region between 0 and $\phi$ in $\Sigma_1$ over  the corresponding, symmetrically placed, opposite region in $\Sigma_2$ will be 
\begin{equation}
\label{eq:7}
\frac {\delta N}{N}=\frac {N_1 - N_2} {N_1 + N_2}= \frac{{\cal D}\sin\phi}{2}=\frac{{\cal D}\cos\theta}{2}.
\end{equation}
Thus the dipole ${\cal D}$ could then be determined from 
${\delta N}/N$ computed for the whole sky ($\phi=\pi/2$). 
\begin{equation}
\label{eq:8}
{\cal D}=\frac {{\cal D}_{\rm o}}{k}=\frac {2}{k}\:\frac {\delta N}{N}.
\end{equation}
where $k$ is a constant, of the order of unity, to be determined numerically for individual samples.

For estimating $k$, Monte--Carlo simulations were made by choosing mock dipoles, oriented in the direction of the CMBR dipole direction, superimposed on randomly assigned source positions in the sky but with a flux-density distribution as of the TGSS sample, so that the source counts remain unaffected. From $\sim 100$ such computer simulations, we estimated $k$ for our TGSS sample for different flux-density cut-offs. 

Our results are presented in Table~IV, which is almost self-explanatory. Dipole ${\cal D}$ was estimated for samples 
containing all sources with flux-density levels $>~\!\!\!S$, starting from $S=250$ mJy and going down to $S=100$ mJy levels. 
Of course the accuracy in our estimate improves as we go to lower flux-density limits since the number of sources increases as $N(>S) \propto S^{-x}$ (with $x\sim 1$).

This method is less prone to statistical errors in the radio source data. For one thing, the errors in the CMBR dipole direction themselves are negligible \cite{1,2,16}, secondly errors in the sky positions of individual sources do not affect the count of total numbers in each hemisphere. Only a very small number of sources in the two small strips of widths $\sim 2$ arcsec (which is the typical error in source positions in the TGSS catalog) on either side of the great circle at the boundary between the two hemispheres could add to the error in dipole magnitude. However the solid angle covered by this strip of width $\sim 2\times2/ (2 \times 10^5) =2\times 10^{-5}$ radian is $\sim 2 \pi\times 2\times 10^{-5}$ sr or $10^{-5}$ fraction of the sky, which for the $N$ values in Table~IV contains only one or two sources, with negligible contribution to $\delta N$ (or even to $\sigma _{\rm N}$) at any flux-density level.

Our estimate of the magnitude of the velocity vector ($v \simeq 3770\pm 340$ km s$^{-1}$) from Table~IV appears order of magnitude higher than the CMBR value ($370$ km s$^{-1}$). The quoted errors for $v$ in Table~IV are from the expected uncertainty $\sigma_N(=\sqrt N)$ in $\delta N(=N_1-N_2)$, the uncertainty here being that of a binomial distribution, similar to that of the random-walk problem (see, e.g. \cite{22}). The corresponding $1\sigma$ uncertainty in $\delta N/N$ is  $1/\sqrt N$, then from Eq.~(\ref{eq:8}), error in $\cal D_{\rm o}$ is $2/\sqrt {N}$.

For a comparison, we also employed the same technique to estimate the magnitude of the velocity vector from the NVSS data, with respect to the CMBR dipole direction. The results for NVSS dataset are also summarized in Table-IV, where $v$ turns out to be $\simeq 1430\pm 290$ km s$^{-1}$. In Table~IV, the four rows of TGSS dataset at various flux-density levels  could be compared to the corresponding four rows of the NVSS dataset. We notice that while the total number $N$ of sources at each flux-density level do match reasonably well (within $1\%$ to  $\stackrel{<}{_{\sim}} 10 \%$), the difference $\delta N$ between two hemispheres in respective flux-density bins, and the thereby derived magnitudes of dipole $\cal D$ and velocity $v$,  differ as much as by a factor of $\sim 2.5$. Of course all estimates of dipole $\cal D$ and velocity $v$ in either dataset are way above the values expected from the CMBR.

From the rms errors ($\sim 300$ to 400 km s$^{-1}$, Table 1V), determined basically by the radio source density field, for the peculiar velocity estimates for the two samples, 
detection of a peculiar velocity like the CMBR value ($\sim 370$ km s$^{-1}$) would not have been possible as it would be within $1\sigma$ level. However, because of the much larger amplitude of the peculiar velocity, by a factor of $\sim 4$ for the NVSS and a factor of $\sim 10$ for the TGSS data, a positive detection became possible at statistically significant, $\sim 4 \sigma$ and $\sim 10 \sigma$ levels, respectively.

\begin{figure}
\includegraphics[width=\columnwidth]{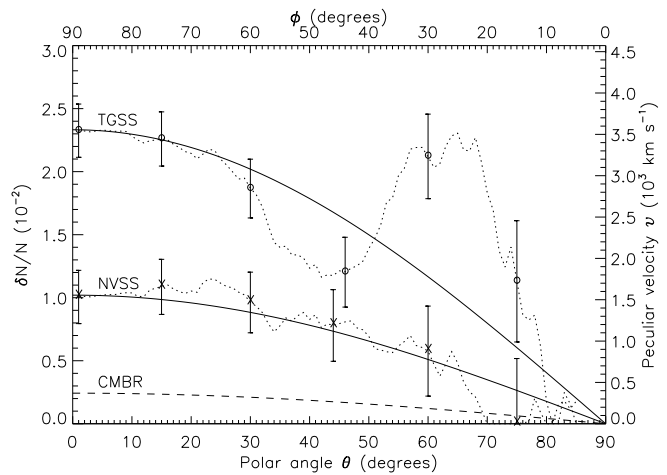}
\caption{A plot of the fractional cumulative excess ${\delta N}/{N}$ against $\phi$, as observed in the sky regions of $\Sigma_1$ over those of $\Sigma_2$, for the TGSS ($>100$ mJy) and NVSS ($>20$ mJy) samples. The corresponding peculiar velocity of the solar system is shown on the right hand scale. 
The dotted lines show the actual observed ${\delta N}/{N}$ values, while the continuous lines show their expected ($\propto \cos \theta$) behavior. Some representative data points are plotted, as circles (o) for the TGSS data and crosses (x) for the NVSS data, with error bars, calculated for a random (binomial) distribution. For a comparison, the dashed line shows ${\delta N}/{N}$, expected for the peculiar velocity equal to the CMBR value, $v=370$ km s$^{-1}$.}
\end{figure}

Here we have explored the radio source dipole by studying any excess of radio source density with respect to the CMBR direction, the latter itself incidentally did not use any information from the radio survey datasets. Moreover, as we move to lower flux-density levels, $\delta N$ seems to steadily increase, specifically, in none of the flux-density bin, for either dataset, we find $N_2$ to be larger than $N_1$. Now if an excess in the sources due to some local clustering in certain regions of the sky were indeed masquerading as a radio source dipole, only in a very contrived situation would one expect to get $N_1 > N_2$ {\em at all flux-density levels} and that too {\em for both radio catalogs}.

From Eq.~(\ref{eq:7}), we expect that the fractional excess, ${\delta N}/{N}$, should have a $\sin\phi$ or $\cos\theta$ dependence. We can verify this  $\sin\phi$ dependence of ${\delta N}/{N}$ by making cumulative counts of $N_1$ and $N_2$ as a function of $\phi$. We should expect large departures from the expected $\sin\phi$ behavior for small $\phi$, not just because of larger statistical fluctuations in a Binomial distribution for small numbers, but also because dipole strength builds up only at larger $\phi$ (number distribution having a dipole ${\cal D}\cos\theta={\cal D}\sin\phi$), i.e., when one approaches the dipole direction at smaller $\theta$. Figure (2) shows the fractional excess, ${\delta N}/{N}$, for both TGSS and NVSS data as a function of $\phi$ or $\theta$. We have used the number counts for the bins $S_{\rm TGSS}>100$ and $S_{\rm NVSS}>20$ as these have the largest number of sources in our samples. As expected, in Fig. (2), we see large fluctuations for smaller $\phi$. While the fractional excess in the NVSS data does stabilize at $\phi \approx 30^\circ$, in the TGSS data it seems to stabilizes only around $\phi \approx 60^\circ$. However, from Fig. (2) it is clear that not only in both TGSS and NVSS cases is the fractional excess, ${\delta N}/{N}$, way above that expected from the CMBR value of peculiar velocity, viz. 370 km s$^{-1}$, but also that  TGSS dipole is much stronger than the NVSS dipole. 

The evidence seems irrefutable that the peculiar velocity of the solar system estimated from the distant radio 
source distributions in sky is indeed much larger than that inferred from the CMBR sky distribution.
Such a statistically significant difference in the estimates of the magnitude of the peculiar velocity is puzzling and one cannot escape the conclusion that there is a genuine disparity in the three reference frames defined by the radio source populations selected at different frequencies and the CMBR. 

From the clustering properties of radio sources in the TGSS angular spectrum on large angular scales, corresponding to multipoles $ 2 \le l \le 30$, the amplitude of the TGSS  angular power spectrum is found to be significantly larger than that of the NVSS, and from that questions have been raised \cite{13} that some unknown systematic errors may be present in the TGSS dataset. At the same time, while the amplitude of the dipole ($l=1$) too is significantly larger than that of the NVSS, the self-consistency of the TGSS dipole in different flux-density bins and the fact that the direction of the dipole coincides with that of the CMBR dipole, indicates that the TGSS dataset may not be affected to such a great extent by systematics.

Now, unless one wants to disregard radio dipoles, derived from TGSS and NVSS datasets, altogether, at this stage one may be left mainly with only two alternatives. One of them is to say that there may be something amiss in the interpretation of the observed dipoles (including the CMBR) as reflecting observer's motion (peculiar velocity of the solar system!) and that the strength of a dipole may not be representing an observer's peculiar speed. In this line of thinking, one will then have to explain the existence of a common direction of all the dipoles and that what is so peculiar about this direction and whether it represents some sort of an ``axis'' of the universe. The other alternative would be to still follow the conventional wisdom that these dipoles are arising as a result of observer's motion and that these dipole magnitudes differing by as much as an order of magnitude, indicates that there may be a large relative motion of the various cosmic reference frames. Either alternative does not fit with the cosmological principle, which is the starting point for the standard modern cosmology. Perhaps it points out to the need for a fresh look at the role of the cosmological principle in the cosmological models.
\section{Conclusions}
From the dipole anisotropy computed for the TGSS dataset, it was found that the dipole strength and the thereby inferred peculiar motion of the solar system is an order of magnitude larger than that inferred from the CMBR dipole. The TGSS dipole is also larger than the NVSS dipole by a factor of $\sim 2.5$. But the direction of the dipole in all these cases turns out to be the same within errors. An obvious inference is that the reference frames determined from some of the most distant observables, viz. the CMBR, the NVSS 1400 MHz dataset of radio sources and the TGSS 150 MHz dataset of radio sources, somehow do not coincide with each other, which raises uncomfortable questions about the cosmological principle, the basis of the modern cosmology.
\section*{Acknowledgements}
Thanks are due to the staff of the GMRT that made the TGSS observations possible. GMRT is run by the National Centre for Radio Astrophysics of the Tata Institute of Fundamental Research. I thank an anonymous referee for his/her comments and suggestions that helped improve the paper.

\end{document}